\documentclass[12pt,preprint]{aastex}

\def\sun{\odot}

\begin{document}
\title{Formation of Kuiper Belt Binaries}
\shortauthors{Schlichting & Sari}
\shorttitle{Formation of Kuiper Belt Binaries }
\author{Hilke E. Schlichting and Re'em Sari}
\affil{California Institute of Technology, MC 130-33, Pasadena, CA 91125} 
\email{hes@astro.caltech.edu, sari@tapir.caltech.edu}

\begin{abstract} The discovery that a substantial fraction of Kuiper Belt
objects (KBOs) exists in binaries with wide separations and roughly equal
masses, has motivated a variety of new theories explaining their
formation. \citet{GLS02} proposed two formation scenarios: In the first, a
transient binary is formed, which becomes bound with the aid of dynamical
friction from the sea of small bodies ($L^2s$ mechanism); in the second, a
binary is formed by three body gravitational deflection ($L^3$
mechanism). Here, we accurately calculate the $L^2s$ and $L^3$ formation rates
for sub-Hill velocities. While the $L^2s$ formation rate is close to previous
order of magnitude estimates, the $L^3$ formation rate is about a factor of 4
smaller. For sub-Hill KBO velocities ($v \ll v_H$) the ratio of the $ L^3 $ to
the $ L^2s $ formation rate is $ 0.05 (v/v_H)$ independent of the small
bodies' velocity dispersion, their surface density or their mutual
collisions. For Super-Hill velocities ($v \gg v_H$) the $L^3$ mechanism
dominates over the $L^2s$ mechanism. Binary formation via the $L^3$ mechanism
competes with binary destruction by passing bodies. Given sufficient time, a
statistical equilibrium abundance of binaries forms.

We show that the frequency of long-lived transient binaries drops
exponentially with the system's lifetime and that such transient
binaries are not important for binary formation via the $L^3$ mechanism,
contrary to \citet{A07}. For the $L^2s$ mechanism we find that the typical
time, transient binaries must last, to form Kuiper Belt binaries (KBBs) for a
given strength of dynamical friction, $D$, increases only logarithmically with
$D$. Longevity of transient binaries (with lifetimes $\geqslant 15
\Omega^{-1}$ as suggested by \citet{A05}) only becomes important for very weak
dynamical friction (i.e. $D \lesssim 0.002$) and is most likely not crucial
for KBB formation.
\end{abstract}


\section{INTRODUCTION} One of the many intriguing discoveries in the Kuiper
Belt is that a substantial fraction of its largest members are binaries. 48
such systems are currently known (for a comprehensive review see
\citet{N07}). Broadly speaking, we can identify two classes of Kuiper Belt
binaries (KBBs). The first class consists of small satellites around the
largest Kuiper Belt objects (KBOs) and the second of roughly equal-mass
binaries with wide separations. The existence of the first class of binaries
is most likely explained by the standard formation scenario involving a
collision and tidal evolution, as has been proposed for the formation of the
Moon and the Pluto-Charon system \citep{HD75,CW76,MK89}. This formation
scenario fails however for the second class of KBBs, since it cannot account
for their wide separations. This has motivated a variety of new theories for
the formation of comparable mass KBBs
\citep[e.g.][]{W02,GLS02,F04,A05,A07}. \citet{W02} proposed a new formation
mechanism for KBBs consisting of a collision between two bodies inside the
Hill sphere of a third. However, in the Kuiper Belt, gravitational scattering
between the two intruders is about 100 times\footnote{For this estimate we
used $\alpha\sim 10^{-4}$ and assumed that the velocity dispersion of the KBOs
at the time of binary formation is less than their Hill velocity, see \S 2 for
details} more common than a collision. Binary formation by three body
gravitational deflection ($L^3$ mechanism), as proposed by \citet{GLS02},
should therefore dominate over such a collisional formation
scenario. \citet{GLS02} proposed a second binary formation scenario: it
consists of the formation of a transient binary, which becomes bound with the
aid of dynamical friction from the sea of small bodies. This is called the
$L^2s$ mechanism. \cite{A05} and \cite{A07} suggest that transient binaries
that spend a long time in their mutual Hill sphere, near a periodic orbit,
form the binaries in the $L^2s$ and $L^3$ mechanism. We address and
investigate the relative importance of these long-lived transient binaries for
the $L^2s$ and $L^3$ formation mechanism and find that they are most likely
not significant for the overall binary formation in the Kuiper Belt. Finally,
\citet{F04} proposed a binary formation mechanism which involves a collision
between two large KBOs which creates a small moon. An exchange reaction
replaces the moon with a massive body with high eccentricity and large
semi-major axis.

In this paper, we accurately calculate the $L^2s$ and $L^3$ formation rates
for sub-Hill KBO velocities and discuss how these rates are modified for
super-Hill velocities. This allows us to determine for which physical
parameters and velocity regime each mechanism dominates the binary
formation. Further, we calculate the frequency of long-lived transient
binaries and assess their importance for the overall KBB formation.

Our paper is structured as follows: In \S 2 we outline our assumptions,
explain our choice of parameters and define variables that will be used
throughout this paper. We calculate the $L^3$ and $L^2s$ formation rates for
sub-Hill KBO velocities in \S 3 and \S 4 respectively. We compare the $L^2s$
and $L^3$ formation rates in the sub-Hill velocity regime in \S 5. In \S 6 we
discuss how these formation rates are modified for super-Hill KBO
velocities. The frequency of long-lived transient binaries and their
significance for the overall KBB formation is calculated in \S 7. Summary and
conclusions follow in \S 8.

\section{DEFINITIONS AND ASSUMPTIONS}
The Hill radius denotes the distance from a KBO at which the tidal forces due
to the Sun and the gravitational force due to the KBO, both acting on a test
particle, are in equilibrium. It is given by
\begin{equation}\label{e1}
R_{H} \equiv a \left( \frac{M}{3 M_{\sun}}\right) ^{1/3} 
\end{equation}
where $a$ is the semi-major axis and $M$ the mass of the KBO. $M_{\sun}$ is
the mass of the sun. We use the `two-group approximation' \citep{GLS02,GLS04}
which consists of the identification of two groups of objects, small ones,
that contain most of the total mass with surface mass density $\sigma$, and
large ones, that contain only a small fraction of the total mass with surface
mass density $\Sigma \ll \sigma $. We assume $\sigma \sim 0.3 \rm{g~cm^{-2}}$
which is the extrapolation of the minimum-mass solar nebular to a heliocentric
distance of $40\rm{AU}$. Estimates from current Kuiper Belt surveys
\citep{TB03,TJL01} yield $\Sigma \sim 3 \times 10^{-4} \rm{g~cm^{-2}}$ for
KBOs with radii of $R \sim 100~\rm{km}$. We use this value of $\Sigma$,
assuming that $\Sigma$ during the formation of KBBs was the same as it is
now. Our choice for $\Sigma$ and $\sigma$ is also consistent with results from
numerical coagulation simulations by \citet{KL99}.

Large bodies grow by the accretion of small bodies. Large KBOs viscously stir
the small bodies, increasing the small bodies' velocity dispersion $u$. As a
result $u$ grows on the same timescale as $R$ provided that mutual collisions
among the small bodies are not yet important.  In this case, $u$ is given by
\begin{equation}\label{e16}
\frac{u}{v_H} \sim \left( \frac{\Sigma}{\sigma \alpha} \right)^{1/2} \sim 3
\end{equation}
where $\alpha = R/R_{H}\sim 10^{-4}$ at $40\rm{AU}$ \citep{GLS02}. $v_H$ is
the Hill velocity of the large bodies which is given by $v_H = \Omega R_H$
where $\Omega$ is the orbital frequency around the sun. The velocity $v$ of
large KBOs increases due to mutual viscous stirring, but is damped by
dynamical friction from the sea of small bodies such that $v < u$. Balancing
the stirring and damping rates of $v$ and substituting for $u$ from equation
\ref{e16}, we find
\begin{equation}\label{e17}
\frac{v}{v_H} \sim \alpha^{-2} \left(\frac{\Sigma}{\sigma}\right)^{3} \sim
0.1.
\end{equation} 
For our choice of parameters, we have that $v < v_{H}$ during the epoch of
formation of bodies with $R\sim 100\rm{km}$. Additionally, we argue that $v$
could not have exceeded $v_{H}$ significantly during satellite formation in
the Kuiper Belt: If $v_{esc} > v > v_{H}$, where $v_{esc}$ is the escape
velocity from the large bodies, then the timescale for mutual collisions is
\begin{eqnarray}\label{e22}
{\tau_{coll}\sim 0.13 \left(\frac {\Sigma}{3 \times 10^{-4}
\rm{g~cm^{-2}}}\right)^{-1} \left(\frac{\rho}{1\rm{g~cm^{-3}}}\right)
\left(\frac{R}{100\rm{km}}\right)\left(\frac{\alpha}{1
\times 10^{-4}}\right)\left(\frac{v}{v_{H}}\right)^{2}  {} } \nonumber\\
{} \left(\frac{\Omega}{7.9 \times
10^{-10}\rm{s^{-1}}}\right)^{-1} \rm{Gyr} {}.
\end{eqnarray}
Equation \ref{e22} shows that the collision timescale among the largest KBOs
($R>100\rm{km}$) would have been excessively long if $v \gg v_{H}$ during
satellite formation. The ubiquity of small satellites around KBOs, that have
radii as large as $\sim 1000\rm{km}$, \citep{BDB06,BS07} and the Pluto-Charon
system \citep{WSM06} suggests that $v < v_{H}$ during their formation, since
their origin is best explained by a giant impact
\citep[e.g.][]{SWS06,BBRS07}. This is supported further by the recent
discovery of a collisional family belonging to $\rm{EL}_{61}$
\citep{BBRS07}. We therefore focus our work on the shear-dominated velocity
regime ($v<v_{H}$). However, we discuss how our results would be modified if
$v>v_H$.

\section{$L^3$ FORMATION RATE}
A transient binary forms when two large KBOs penetrate each other's Hill
sphere. This transient binary must lose energy in order to become
gravitationally bound. In the $L^3$ mechanism the excess energy is carried
away by an encounter with a third massive body. We calculate the binary
formation rate via the $L^3$ mechanism in the shear-dominated velocity
regime. Since the growth of inclinations is suppressed in the shear-dominated
velocity regime the disk of KBOs is effectively two-dimensional
\citep{WS1993,R2003,GLS04}. We therefore restrict this calculation to two
dimensions. As initial condition, we assume that all bodies are on circular
orbits. We chose to work in the rotating frame with the $x$-axis pointing
radially outwards and the $y$-axis in the prograde direction. For a
gravitational deflection of three equal-mass bodies, the $L^3$ formation rate
per body is
\begin{equation}\label{e19}
FR_{L^3}=\int_{\gamma=-\infty}^{\infty}  \int_{b_2>b_1}^{\infty}
\int_{b_1=0}^{\infty}\left(\frac{\Sigma}{\frac{4 \pi}{3}\rho R^{3}}\right)^{2}
\frac{3}{2} b_1 \Omega F_{L^3}(b_1,b_2,\gamma) db_1 db_2 d\gamma.
\end{equation}
$\Sigma/(4\pi \rho R^3/3)$ is the surface number density of the KBOs. $b_1$
and $b_2$ are the relative initial separations in the $x$-direction between
bodies 1 and 2 and bodies 1 and 3 respectively. $\gamma$ is the offset in the
$y$-direction body 3 would have when bodies 1 and 2 would encounter each other
had their relative velocity been solely due to the Kepler shear of the disk:
$3 b_1 \Omega /2$. Finally, $F_{L^3}(b_1,b_2,\gamma)$ is a function that takes
on the value 1 if the encounter resulted in the formation of a binary between
any two of the three KBOs involved and 0 otherwise. The choice of limits on
the integrals in equation \ref{e19} ensures no double counting of the
binaries. Expression (\ref{e19}) can be written as
\begin{equation}\label{e20} 
FR_{L^3}=A_{L^3} \left( \frac {\Sigma}{\rho R} \right)^2
\alpha^{-4}  \Omega 
\end{equation} 
where 
\begin{equation}\label{e4} 
A_{L^3}=\left( \frac{27}{32 \pi^2} \right) \int_{\gamma=-\infty}^{\infty}
\int_{b_2>b_1}^{\infty} \int_{b_1=0}^{\infty} F_{L^3}(b_1,b_2,\gamma)
\left(\frac{b_1}{R_{H}}\right) \left( \frac{db_1}{R_{H}}\right) \left(
\frac{db_2}{R_{H}} \right) \left(\frac{d\gamma}{R_H}\right).
\end{equation} 
Expression (\ref{e20}) agrees with the order of magnitude estimate of
\citet{GLS02} if we set $A_{L^3}=1$.  It is the value of the constant
$A_{L^3}$ we determine here. Since we are interested in close encounters
among the KBOs, their interaction is well described by Hill's equations
\citep{H78,GT80,PH86} that we modify to include three equal mass bodies
besides the Sun. The equations of motion, with length scaled by $R_{H}$ and
time by $\Omega^{-1}$, for body 1 are given by
\begin{equation}\label{e2}
\ddot{x}_{1}-2
\dot{y}_{1} -3x_{1}=-\frac{3 (x_{1}-x_{2})}
{((x_{1}-x_{2})^2+(y_{1}-y_{2})^2)^{3/2}}-\frac{3 (x_{1}-x_{3})}
{((x_{1}-x_{3})^2+(y_{1}-y_{3})^2)^{3/2}} \end{equation}
\begin{equation}\label{e3} 
\ddot{y}_{1}+2\dot{x}_{1}=-\frac{3
(y_{1}-y_2)} {((x_{1}-x_2)^2+(y_{1}-y_2)^2)^{3/2}}-\frac{3 (y_{1}-y_{3})}
{((x_{1}-x_{3})^2+(y_{1}-y_{3})^2)^{3/2}}.
\end{equation}
The subscripts 1, 2 and 3 label the $x$- and $y$-coordinates of KBO 1, 2, and
3 respectively. Similar equations of motion can be obtained for bodies 2 and
3. $F_{L^3}(b_1,b_2,\gamma)$ is calculated by numerically integrating the
equations of motion. A binary formation event is detected in the following
way: The equations of motion of the three bodies are integrated until a time
that corresponds to a separation of at least $30 R_H$ between all three bodies
(after their conjunction), assuming that their relative velocity is solely due
to their Keplerian sheer (i.e. ignoring the actual gravitational interaction
between the bodies), plus an additional time of $120 \Omega^{-1}$. If after
this time the separation between two bodies is still less than $3R_H$ a binary
is considered to have formed. We chose a separation of $3 R_H$ instead of
$R_H$ to allow for binary orbits that reach slightly outside $R_H$. Numerical
integrations are terminated early if the separation between KBOs becomes less
than $10^{-4} R_H$ and these events are not counted towards the binaries
formed. This serves two purposes: first of all, $10^{-4} R_H$ roughly
corresponds to the separation at which physical collisions occur in the Kuiper
Belt. Secondly, by introducing a minimum separation, we prevent divergence in
the equations of motion. This cut-off limits, strictly speaking, the validity
of the value of $A_L^3$ calculated here to binary formation at heliocentric
distances of $\sim 40\rm{AU}$ since the separation in units of $R_H$,
corresponding to collisions among the KBOs, is inversely proportional to the
heliocentric distance. In order to determine $A_{L^3}$ we need to cover the
three dimensional parameter space spanned by $b_1$, $b_2$ and $\gamma$. We
chose a spacing of $0.1 R_{H}$ for all three parameters. $12.5 R_H$ is chosen
as the upper limit for $b_1$ and $b_2$, the upper limit for $|\gamma|$ is $25
R_H$. The given limits and resolution require numerical integrations of $\sim
4 \times 10^6$ orbits. We obtain
\begin{equation}\label{e5} 
A_{L^3}=0.28 \pm 0.01
\end{equation} 
where $0.01$ is the estimated Poisson error. We repeated the calculation for
$A_{L^3}$ with randomly chosen grid points for $b_1$, $b_2$ and $\gamma$ and
the same number of numerical integrations and confirmed that the value of
$A_{L^3}$ is insensitive to the grid points chosen. $A_{L^3}$ tends
to $0.35$ in the limit that the bodies are treated as point masses
(i.e. the limit that the cut-off tends to zero). We will use $A_{L^3}=0.28$
since it corresponds to the physically relevant situation in the Kuiper
Belt. This yields a binary formation rate of
\begin{eqnarray}\label{e6}
{FR_{L^3}=(6.3 \pm 0.2) \times 10^{-8} \left(\frac {\Sigma}{3 \times 10^{-4}
\rm{g~cm^{-2}}}\right)^2 \left(\frac{\rho}{1\rm{g~cm^{-3}}}\right)^{-2}
  \left(\frac{R}{100\rm{km}}\right)^{-2} {} } \nonumber\\
{} \left(\frac{\alpha}{1 \times 10^{-4}}\right)^{-4}
\left(\frac{\Omega}{7.9 \times 10^{-10}\rm{s^{-1}}}\right) \rm{yr^{-1}} {},
\end{eqnarray} 
which is smaller by $1/A_{L^3} \sim 4$ than the order of magnitude estimate of
\citet{GLS02}.

\section{$L^2s$ FORMATION RATE}
\begin{figure}
\plotone{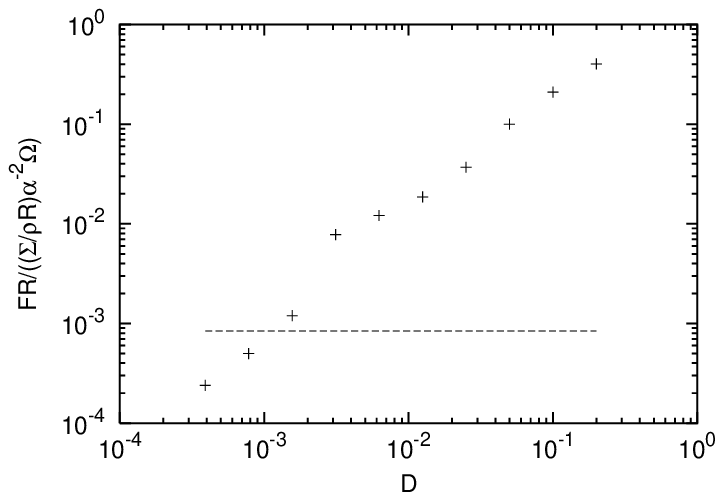} 
\caption{Binary formation rate as a function of dynamical friction strength $
  D $. The crosses correspond to the formation rate via the $ L^2s $
  mechanism and the dashed horizontal line corresponds to the $ L^3 $
  formation rate for $ (\Sigma/\rho R)\alpha^{-2} = 3 \times 10^{-3}$. The $
  L^2s $ formation rate is proportional to $D$. In \S 4 we estimate that
  $D\sim 0.12$ as a result of which $ FR_{L^3}/FR_{L^2s} \sim 0.005 $.
}
\label{fig1} 
\end{figure}

So far, we have only considered binary formation due to an encounter with a
third body that carries away the excess energy. However, binary formation
might also occur due to dynamical friction generated by the sea of small
bodies ($L^2s$ mechanism). The random velocity of large KBOs is damped due to
gravitational interactions with many small bodies. Since it is not feasible to
examine the interactions with each small body individually, their net effect
is modeled by an averaged force which acts to damp the large KBOs'
non-circular velocity. We parameterize the strength of the damping by a
dimensionless quantity $D$ defined as the fractional decrease in non-circular
velocity due to dynamical friction over a time $\Omega^{-1}$:
\begin{equation}\label{e41}
D\sim \frac{ \sigma}{\rho R} \left(\frac{u}{v_H}\right)^{-4} \alpha^{-2} \sim
\frac{\Sigma}{\rho R} \alpha^{-2} \left(\frac{v}{v_H}\right)^{-1}.
\end{equation} 
The first expression is simply an estimate of dynamical friction by the sea of
small bodies assuming $u>v_H$. The second expression describes the mutual
excitation among the large KBOs for $v<v_H$. These two expressions can be
equated since the stirring among the large KBOs is balanced by the damping due
to dynamical friction. In fact, if $v$ is defined as the product of the median
eccentricity and the orbital velocity, we can calculate the exact relationship
between $D$ and $(v/v_H)$ since the velocity distribution in the
shear-dominated velocity regime has been fully determined (see
\citet{CS06,CSS07}). Defining $v$ as the product of the median eccentricity
and the orbital velocity, we obtain
\begin{equation}\label{e51}
D=4.1 \frac{\Sigma}{\rho R} \alpha^{-2} \left(\frac{v}{v_H}\right)^{-1}.
\end{equation} 
For $\rho \sim 1 \rm{g~cm^{-3}}$ and our estimates for $(v/v_H)$, $\Sigma$ and
$R$ from \S 2 we find $D \sim 0.12$. We calculate the binary formation rate
for equal mass bodies via the $L^2s$ mechanism in shear-dominated velocity
regime. As in \S 3, we restrict this calculation to two dimensions with
circular motion as initial conditions for the large KBOs and use the same
coordinate system as in \S 3. The binary formation rate per body via the
$L^2s$ mechanism can be written as
\begin{equation}\label{e199}
FR_{L^2s}=\int_{b=0}^{\infty}\left(\frac{\Sigma}{\frac{4 \pi}{3}\rho
R^{3}}\right) \frac{3}{2} b \Omega F_{L^2s}(D,b) db
\end{equation}
where $\Sigma/(4\pi \rho R^3/3)$ is the surface number density of the KBOs and
$b$ is the relative initial separation in the $x$-direction between the two
KBOs. $F_{L^2s}(D,b)$ is a function that takes on the value 1 if the
encounter resulted in the formation of a binary for a given $D$ and $b$ and 0
otherwise. Equation \ref{e199} can be written as
\begin{equation}\label{e9} 
FR_{L^2s}=A_{L^2s} D \left(\frac{\Sigma}{\rho R}\right) \alpha^{-2} \Omega
\end{equation} 
where
\begin{equation}
A_{L^2s}=D^{-1}\left(\frac{9}{8 \pi}\right) \int_{b=0}^{\infty} F_{L^2s}(D,b)
\left(\frac{b}{R_H}\right) \left(\frac{db}{R_H}\right).
\end{equation} 
\citet{GLS02} showed, using numerical integrations, that $FR_{L^2s}$ is indeed
proportional to $D$. Here we want to determine the actual value of
$A_{L^2s}$. In Hill coordinates the equations of motion of the two KBOs can be
decomposed into their center of mass motion and their relative motion with
respect to one another. The relative motion of two equal mass KBOs, including
the dynamical friction term, is governed by
\begin{equation}\label{e10} 
\ddot{x}-2 \dot{y}
-3x=-\frac{6 x}{(x^2+y^2)^{3/2}}-D \dot{x}
\end{equation} 
\begin{equation}\label{e11} 
\ddot{y}+2
\dot{x}=-\frac{6 y}{(x^2+y^2)^{3/2}} -D (\dot{y}+1.5x).
\end{equation} 
where $x$ and $y$ correspond to the relative separation between the two KBOs
in the $x$ and $y$ direction respectively. Again, length has been scaled by
$R_{H}$ and time by $\Omega^{-1}$. Equations \ref{e10} and \ref{e11} are
integrated for different values of $D$ and impact parameters ranging from $2.2
R_H$ to $3.2 R_H$. Impact parameters outside this range result in a distance
of closest approach between the two KBOs of more than $R_H$. Figure \ref{fig1}
shows that the rate of binary formation is proportional to $D$. The value of
$A_{L^2s}$, estimated from the line of best fit, is 1.4. This yields a binary
formation rate of
\begin{eqnarray}\label{e12} 
{FR_{L^2s}= 1.3 \times 10^{-5} \left(\frac{D}{0.12}\right) \left(\frac
  {\Sigma}{3 \times 10^{-4}\rm{g~cm^{-2}}}\right)
  \left(\frac{\rho}{1\rm{g~cm^{-3}}}\right)^{-1}
  \left(\frac{R}{100\rm{km}}\right)^{-1} {} } \nonumber\\ {}
  \left(\frac{\alpha}{1 \times 10^{-4}}\right)^{-2} \left(\frac{\Omega}{7.9
  \times 10^{-10}\rm{s^{-1}}}\right) \rm{yr^{-1}} {}.
\end{eqnarray}
Using equation \ref{e41} we can retrieve the scalings of
\citet{GLS02}. Although, we know the exact $L^2s$ formation rate for
a given $D$ and have an exact expression for $D$ in terms of $v$ (see equation
\ref{e51}), the relation between $v$ and the actual physical parameters,
i.e. the numerical coefficient in equation \ref{e17}, which is needed for a
precise value of $D$ is uncertain to a factor of order unity.

Contrary to claims by \citet{A05} and \citet{A07}, the $L^2s$ mechanism does
predict a mass-ratio selection. This can be seen from the first part of
equation \ref{e41}. For a given $u$, we have that $D \propto R^3$ since
$v_H\alpha^{-1/2} \sim v_{esc}\propto R$. Large KBOs experience stronger
dynamical friction (larger $D$) than smaller ones. This is not at all
surprising and is a general feature of dynamical friction \citep{C43,BT87}. We
can write $D=D_0M$ where $D_0\sim \sigma G^2/u^4$. $D_0$ is a constant
independent of the KBO mass for a given $\sigma$ and $u$. For two KBOs with
masses $M_1$ and $M_2$, the position of body 1 essentially coincides with the
center of mass of the two bodies provided that $M_1/M_2 \gg 1$. In the limit
that the KBOs random velocity tends to zero and that $D_0M_2 \ll 1$ we can
place body 1 at the origin of the Hill coordinate system and treat the center
of mass as stationary throughout the interaction. In this limit we find
that the relative motion of the two KBOs is governed by
\begin{equation}\label{e777}
\ddot{x}-2 \dot{y} -3x=-\frac{3x}{(x^2+y^2)^{3/2}}-2D_0M_2 \dot{x}
\end{equation} 
\begin{equation}\label{e778}
\ddot{y}+2 \dot{x}=-\frac{3y}{(x^2+y^2)^{3/2}} -2D_0M_2 (\dot{y}+1.5x)
\end{equation}
where length is scaled by $R_H$ of KBO 1 and time is scaled by $\Omega^{-1}$.
For extreme mass-ratio binaries the relevant strength of the dynamical
friction that enters equations \ref{e777} and \ref{e778} is twice that acting
on the small body (i.e. $2D_0M_2$), and significantly less than that acting on
the large body (i.e. $D_0M_1$). The $L^2s$ mechanism therfore favors the
formation of comparable mass binaries from the largest available bodies over
high mass-ratio ones. It is an open question, whether this preference for
comparable mass binaries remains after the Kuiper Belt mass spectrum during
their formation and their survival probability are accounted for.

\section{COMPARISON OF $L^2s$ AND $L^3$ FORMATION RATES}
We are now able to compare the binary formation rates for the $L^2s$ and
$L^3$ mechanism for sub-Hill velocities.
The ratio of the $L^3$ to $L^2s$ formation rates is
\begin{equation}\label{e13}
\frac{FR_{L^3}}{FR_{L^2s}}=0.20 D^{-1} \frac{\Sigma}{\rho R} \alpha^{-2}
=0.05\frac{v}{v_H}
\end{equation}
where we substituted for $D$ using the exact relationship from equation
 \ref{e51}. It is remarkable that this expression depends explicity only
 on $v/v_H$ and is independent of what sets $D$. It is therefore independent
 of the velocity dispersion of the small bodies, their surface density or the
 importance of collisions among the small bodies. We therefore conclude that
 for $v\ll v_H$, binaries in the Kuiper Belt formed primarily due to dynamical
 friction rather than three body encounters. Figure \ref{fig1} shows the
 $L^2s$ and $L^3$ formation rates as a function of $D$. For our estimate of
 $(v/v_H)\sim 0.1$, we have that $ FR_{L^3}/FR_{L^2s} \sim 0.005 $.

\section{Super-Hill Velocity: $v>v_H$}
Obviously there is some uncertainty in what the actual values of $\sigma$ and
$\Sigma$ were during binary formation. For a few times larger value of $
\Sigma $ with $ \sigma $ unchanged, we enter the regime in which $v>v_H$ (this
can be seen from equation \ref{e16}). Although it
is rather unlikely that $v \gg v_H$ during binary formation (see \S 2) we
discuss here briefly how this would affect the $L^2s$ and $L^3$ formation
rates.

For $v>v_H$ the velocity dispersion of the large bodies is still set by the
balance between their mutual stirring and the damping due to dynamical
friction generated by the sea of small bodies. Therefore, dynamical friction
shrinks the orbit of a KBB with a mutual orbital velocity $v_B$ at a rate
\begin{equation}\label{e42}
D\Omega\sim \frac{\Sigma}{\rho R} \alpha^{-2} \Omega
\left(\frac{v}{v_H}\right)^{-4}
\end{equation}
where we assume that $v_B<u$. For $v_B \lesssim v$, binaries are broken up by
passing KBOs at a rate
\begin{equation}\label{e101}
R_{break}\sim \frac{\Sigma }{\rho R} \alpha^{-2} \Omega
\left(\frac{v}{v_H}\right)^{-2}\left(\frac{v_B}{v_H}\right)^{-2}.
\end{equation}
The ratio of these two rates yields
\begin{equation}\label{e600}
\frac{D\Omega}{R_{break}}\sim \left(\frac{v_B}{v}\right)^{2}.
\end{equation}
Since the ratio in equation \ref{e600} is $<1$ for $v_B<v$, we conclude that
KBBs with separations $R_B> R_H (v_H/v)^2$ (i.e. KBBs with $v_B < v$) tend to
be broken up by passing KBOs. Binaries with separations of $R_{crit}= R_H
(v_H/v)^2$ or less, tend to survive. The cross section for the $L^3$ mechanism
is therefore reduced with respect to the sub-Hill case. The probability of
having a KBO within $R_{crit}$ of a given KBO is $(\Sigma \Omega)/(\rho R^3
v)~ R_{crit}^3$ where $(\Sigma \Omega)/(\rho R^3 v)$ is the volume number
density of KBOs. The flux of KBOs into area $R_{crit}^2$ is $(\Sigma
\Omega)/(\rho R^3 v)~v R_{crit}^2$. The super-Hill formation rate for tight
binaries with separations $\sim R_{crit}$, via the $L^3$ mechanism, is
therefore
\begin{equation}\label{e167}
FR_{L^3}\sim \left( \frac {\Sigma \Omega}{\rho R^3 v} \right)^2 v R_{crit}^5
\sim \left( \frac {\Sigma}{\rho R} \right)^2 \alpha^{-4}
\left(\frac{v_H}{v}\right)^{11} \Omega
\end{equation}
(see also \citet{N07}). In addition to tight binaries with separations of
 $R_{crit}$ and less, there exist a second class of binaries with larger
 separations. Binaries with separations $R_B>R_{crit}$ are constantly created
 and destroyed via the $L^3$ mechanism. KBBs can form from two KBOs that
 approach each other with relative velocity $v_B \lesssim v$ while a third KBO
 removes energy, through gravitational scattering, enabling the KBOs to get
 bound. Since we are selecting bodies with relative velocities $\sim v_B$ or
 less, the number of KBOs that can form binaries with separation $R_B= R_H
 (v_H/v_B)^2$ is reduced by $\sim (v_B/v)^3$. The formation rate for binaries
 with separation $R_B=R_H (v_H/v_B)^2$ is
\begin{equation}\label{e169}
FR_{L^3}(R_B>R_{crit})\sim \left( \frac {\Sigma}{\rho R} \right)^2
\alpha^{-4} \left(\frac{v_H}{v}\right)^6 \left(\frac{v_H}{v_B}\right)^{5}
\Omega.
\end{equation}
These wider binaries ($R_B>R_{crit}$) have a higher formation rate compared to
the tight ones which have a separation $\sim R_{crit}$. The ratio of the
formation rate (equation \ref{e169}) to the destruction rate (equation
\ref{e101}) yields an equilibrium abundance of binaries per KBO at any given
time that is given by
\begin{equation}\label{e170}
\frac{N_{KBB}}{N_{KBO}}\sim \frac{\Sigma}{\rho R} \alpha^{-2}
\left(\frac{v_H}{v}\right)^4 \left(\frac{v_H}{v_B}\right)^{3}.
\end{equation}
The number of binaries scales as $(R_B/R_H)^{3/2}$. Binaries with separation
$R_B$ are therefore $(R_B/R_{crit})^{3/2}\sim(v/v_B)^3$ times more common than
those with separation $R_{crit}$ provided there is sufficient time for the
equilibrium to be established. The same statistical equilibrium abundance can
be derived using phase space arguments. The phase space number density of KBOs
is $(\Sigma \Omega)/(\rho R^3 v^4)$. The phase space volume corresponding to a
binary separation $R_B$ and velocity $v_B$ is $R_B^3 v_B^3=R_H^3 v_H^3
(v_H/v_B)^3$. Multiplying the KBO phase space number density by the binary
phase space volume yields a statistical equilibrium abundance per KBO of
\begin{equation}
\frac{N_{KBB}}{N_{KBO}}\sim \frac{\Sigma \Omega}{\rho R^3 v^4} R_H^3 v_H^3
\left(\frac{v_H}{v_B}\right)^3\sim \frac{\Sigma}{\rho R} \alpha^{-2}
\left(\frac{v_H}{v}\right)^4 \left(\frac{v_H}{v_B}\right)^{3},
\end{equation}
which is in agreement with the binary abundance derived in
(\ref{e170}). Whether any of these binaries would survive the dynamical
excitation of the Kuiper Belt remains an open question.

The $L^2s$ mechanism fails in creating binaries with separations $\sim
R_{crit}$ since dynamical friction is not able to dissipate sufficient energy
for tight binaries to form. Dynamical friction is only able to assist in
the formation of binaries with wide separations ($\sim R_H$) that form from
KBOs that happen to approach each other with low relative velocities ($\sim
v_H$). This reduces the number density of KBOs that can participate in
binary formation by a factor of $\sim (v_H/v)^{3}$. In this case, the $L^2s$
formation rate is given by
\begin{equation}\label{e201}
FR_{L^2s}(R_B\sim R_H)\sim D \left(\frac{\Sigma}{\rho R}\right) \alpha^{-2}
\left(\frac{v_H}{v}\right)^4\Omega \sim \left(\frac{\Sigma}{\rho R}\right)^2
\alpha^{-4} \left(\frac{v_H}{v}\right)^8\Omega
\end{equation}
where  we  have  substituted for  $D$  from  equation  \ref{e42} in  the  last
step. These wide binaries  face the same fate as the wide  ones formed via the
$L^3$ mechanism  in that they will be  broken up quickly due  to scattering of
other large  bodies. However,  the $L^2s$ mechanism  does not  even contribute
significantly  to  the binary  equilibrium  abundance  calculated in  equation
\ref{e170} since $FR_{L^2s}(R_B\sim R_H)/FR_{L^3}(R_B\sim R_{H})\sim (v_H/v)^2
\ll 1$. Therfore, the $L^2s$ mechanism does not play an important role in KBB
formation if super-Hill velocities prevail.

In summary, the $L^3$ mechanism forms tight binaries, that tend to be saved
from break up, at a rate that is reduced by a factor of $(v_H/v)^{11}$
compared to the sub-Hill case. In addition, the $L^3$ mechanism forms wider
binaries $(R_B>R_{crit})$, at a higher rate that is 'only' reduced by a factor
of $(v_H/v)^{6}(v_H/v_B)^5$ relative to the sub-Hill rate. These wide binaries
are constantly created and destroyed leading to an equilibrium abundance of
binaries that scales as $(R_B/R_H)^{3/2}$. The $L^2s$ mechanism is not
important if KBOs have super-Hill velocities.

\section{FREQUENCY OF LONG-LIVED TRANSIENT BINARIES AND THEIR SIGNIFICANCE FOR
  BINARY FORMATION} \citet{A05} propose that transient binaries, that
spent a time of $15 \Omega^{-1}$ ($\sim 600 \rm{yr}$ at $40\rm{AU}$) or longer
in their mutual Hill sphere, near a periodic orbit, are responsible for binary
formation in the $L^2s$ and $L^3$ mechanism. Here, we determine how the
frequency of long-lived transient binaries depends on the transient binary
lifetime. This allows us to quantify the importance of long-lived transient
binaries for the overall binary formation. Finally, we address the
significance of long-lived transient binaries for the $L^2s$ and $L^3$
formation mechanism.

\subsection{Frequency of long-lived Transient Binaries} 
\begin{figure}
\plotone{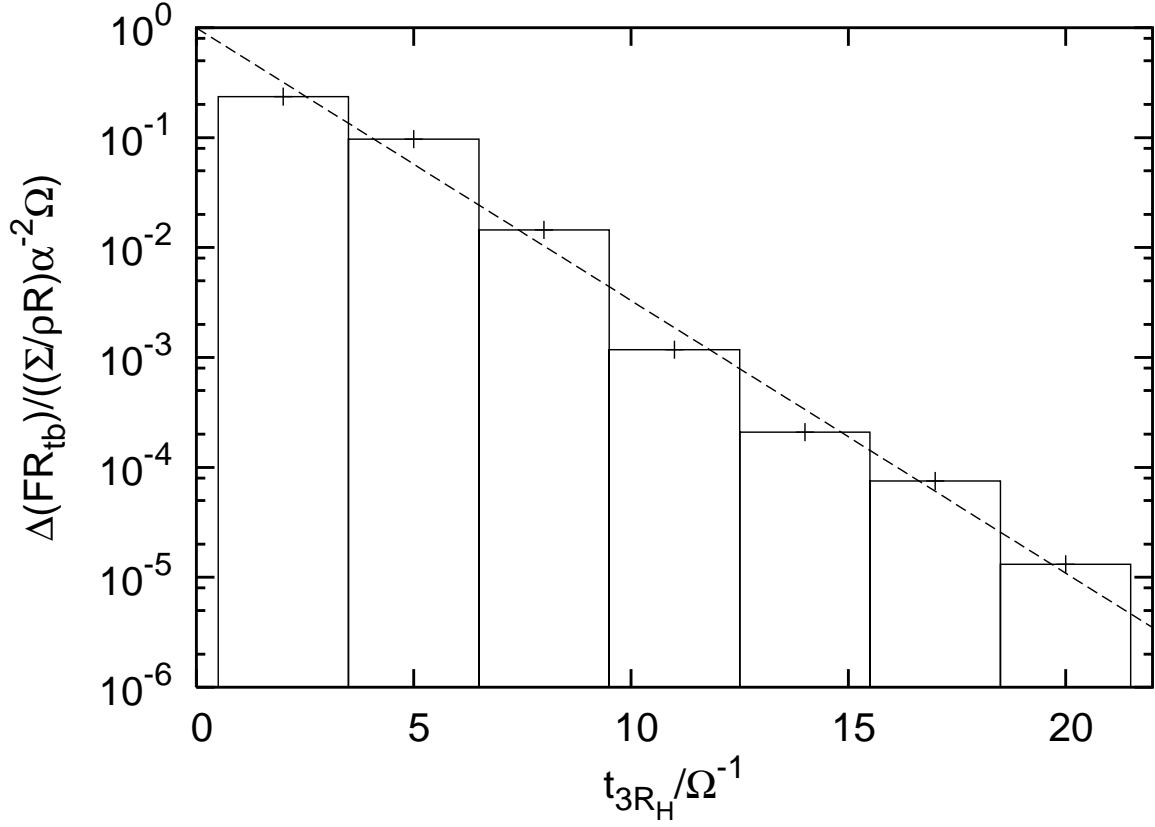} 
\caption{Differential transient binary frequency $\Delta (FR_{tb})$ as a
  function of the transient binary lifetime $t_{3R_H}$ in the shear dominated
  velocity regime. $t_{3R_H}$ is the time the transient binary separation is
  less than $3 R_H$. The frequency decreases exponentially with $
  t_{3R_H}\Omega$.  
}
\label{fig2} 
\end{figure}

First, we assess how common long lived transient binaries are. We integrate
equations \ref{e10} and \ref{e11} without the dynamical friction term and
determine the time $t_{3R_{H}}$ over which the separation between the two KBOs
is less than $3 R_{H}$ for all KBOs that approach one another to $R_H$ and
less. We chose to calculate the time the two KBOs spent with a separation of
less than $3R_{H}$ to allow for orbits that reach slightly outside of $R_{H}$
but return back to within $R_{H}$ during the encounter. We integrate $\sim
10^5$ orbits in total with impact parameters ranging from $2.2 R_{H}$ to $3.2
R_{H}$. Impact parameters outside this range result in a distance of closest
approach between the two KBOs of more than $R_H$. As initial conditions, we
assume that the orbits of the bodies are circular. Figure \ref{fig2} shows
that the frequency of transient binaries decreases exponentially with the
transient binary lifetime, $t_{3R_H}$. The line of best fit yields a
differential transient binary frequency, valid for $t_{3R_H}\gtrsim 1
\Omega^{-1}$, of
\begin{equation}\label{e14} 
\frac{d (FR_{tb})}{d(t_{3R_H} \Omega)}= 1.0 \times 10^{-(0.25 t_{3R_H}
  \Omega)}\frac{\Sigma}{\rho R} \alpha^{-2} \Omega.
\end{equation} 
The frequency of transient binaries that spend a time of $ \gtrsim 15
\Omega^{-1}$ with a separation of less than $3R_H$ is 3 orders of magnitude
smaller than that of short-lived ones with $t_{3R_H} \gtrsim 3 \Omega
^{-1}$. The analysis discussed here was carried out assuming that the KBOs
approach each other with relative velocities $v_{rel} < v_H$. Long-lived
transient binaries do not exist for bodies that encounter each other at
$v_{rel} \gg v_H$. This can be understood by looking at the Jacobi
constant. The Jacobi constant in Hill coordinates with length scaled by $R_H$
and time by $\Omega^{-1}$ is given by
\begin{equation}\label{500}
C_J=3x^2-z^2 +\frac{12}{(x^2+y^2+z^2)^{1/2}} -\dot{x}^2 -\dot{y}^2 -\dot{z}^2.
\end{equation}  
KBOs that approach each other with $v_{rel}\gg v_H$ at $R_H$ will experience
at most one encounter before they separate. Evaluation of their Jacobi
constant at $R_H$ yields that it is large and negative. In order to experience
multiple encounters, KBOs must approach each other with $v_{rel}\sim v_H$ at
$R_H$ which corresponds to $ C_J$ of order unity. Since the Jacobi constant is
a conserved quantity, we can be sure that no long-lived transient binaries
exist for KBOs that encounter each other at $v_{rel}\gg v_H$. Long-lived
transient binaries therefore offer no solution to the fine tuning problem,
contrary to claims by \cite{A07}. For KBOs with a given velocity distribution
there always exist a few bodies that have $v_{rel}<v_H$ even if $v\gg
v_H$. Such bodies can give rise to long-lived transient binaries in the
same way that they can form wide binaries (see \S 6 for details), but the
frequency of transient binaries due to such bodies is reduced by a factor of
$(v_H/v)^4$.

\subsection{Importance of long-lived Transient Binaries in the $L^3$ Formation Mechanism} 
\cite{A07} claim that the probability of binary formation from transient
binaries with $t_{3R_H}\lesssim 2.5 \Omega^{-1}$ is extremely small and they
therefore include only transient-binaries with $t_{3R_H}\gtrsim 5 \Omega^{-1}$
in the main set of their integrations. However, their conclusion, that the
probability of binary formation from transient binaries with $t_{3R_H}\lesssim
2.5 \Omega^{-1}$ is extremely small, is due to a bias in their initial
conditions that discriminates against binary formation from transient binaries
with $t_{3R_H}\lesssim 5 \Omega^{-1}$. The shortcoming of their analysis is
due to the fact that they launch the third body from an initial separation $>
38 R_H$ when the first two KBOs come within a few $R_H$ of each
other\footnote{The numerical values stated by \citet{A07} are multiplied by a
factor of $2^{1/3}$ to compensate for the different definitions of $R_H$
}. Since \citet{A07} select the initial conditions for the third body such
that it penetrates within $2.5 R_H$, the largest impact parameter is $\sim 4.5
R_H$. The minimum time it takes for the third body to come within a few $R_H$
of the transient binary is therefore $ \sim 38 R_H/(1.5 \times 4.5 R_H \Omega)
\sim 5.6 \Omega^{-1}$. The third body therefore only reaches the vicinity of
the transient binary for $t_{3R_H}\gtrsim 5.6 \Omega^{-1}$, but it is exactly
this proximity of the third body that is required for binary formation by
strong gravitational scattering. This explains why \citet{A07} find such a
small probability for binary formation by transient binaries with
$t_{3R_H}\lesssim 5.6 \Omega^{-1}$. The range of impact parameters that lead
to binary formation is comparable for short- and long-lived transient
binaries. This means that the transient binary lifetime is the only advantage
long-lived transient binaries have compared to short-lived ones, in terms of
binary formation likelihood. However, as we show in \S 7.1, the frequency of
transient binaries drops exponentially as a function of their lifetime. The
ratio of the binary formation rate due to short-lived transient binaries
($t_{3R_H}\gtrsim 3 \Omega^{-1}$) compared to that due to long-lived ones
($t_{3R_H}\gtrsim 15 \Omega^{-1}$), is therefore
\begin{equation}
\frac{FR(t_{3R_H}\gtrsim 3 \Omega^{-1})}{FR(t_{3R_H}\gtrsim 15
\Omega^{-1})}\sim \frac{3 \Omega^{-1}}{15 \Omega^{-1}} \frac{10^{-(0.25 \times
3)}}{10^{-(0.25 \times 15)}} \sim 200.
\end{equation} 
Although the binary formation rate scales linearly with transient-binary
lifetime, the frequency of transient binaries drops exponentially as a
function of its lifetime. Therefore, long-lived transient binaries are not
important for binary formation via the $L^3$ mechanism.

\subsection{Importance of long-lived Transient Binaries in the $L^2s$ Formation Mechanism} 
\begin{figure} 
\plotone{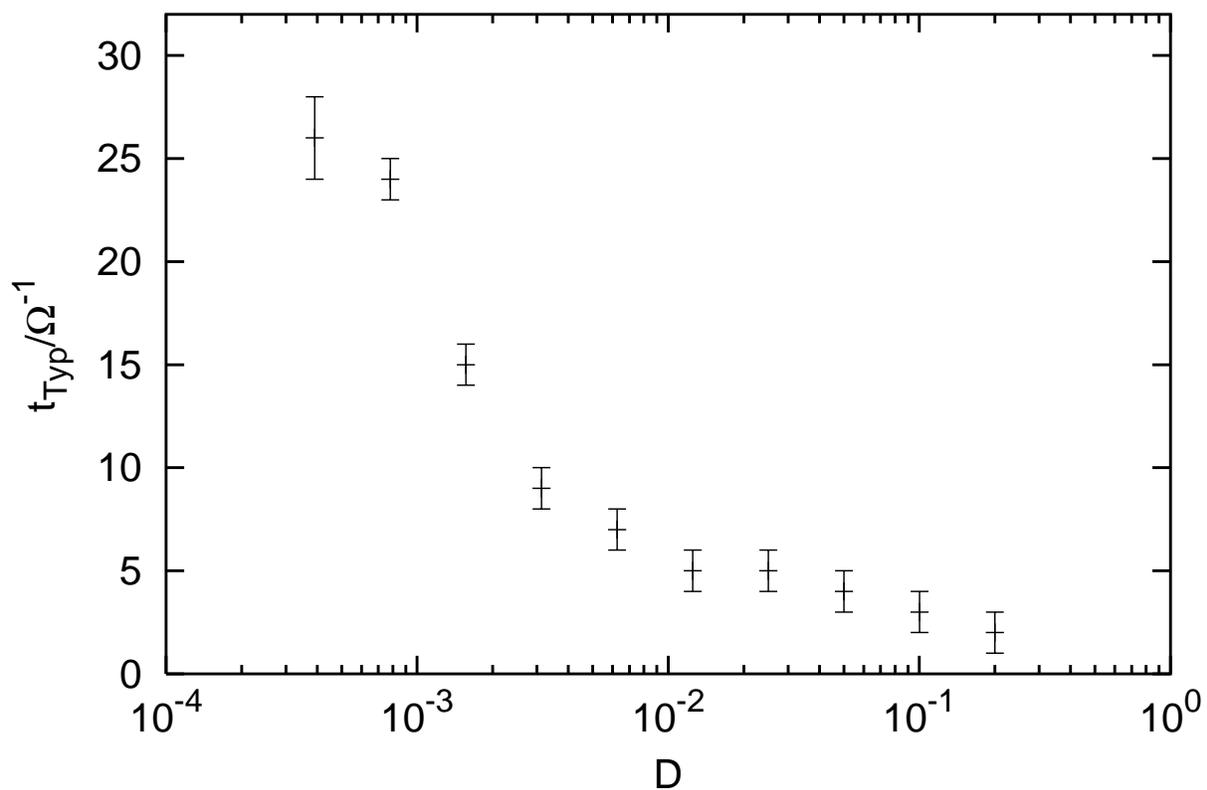} 
\caption{The variation of $t_{Typ}$, the time taken for 50\% of binaries to
  get bound, plotted against strength of dynamical friction $D$. For about two
  orders of magnitude change in $D$ ($D \sim 0.2$ to $D \sim 0.002$) $t_{Typ}$
  only changes from $\sim 2 \Omega^{-1}$ to $\sim 10 \Omega^{-1}$. A rapid rise
  in $t_{Typ}$ occurs for $D \lesssim 0.001$.  
}
\label{fig3} 
\end{figure}

In general, KBOs that spend a longer time in the Hill sphere, lose more energy
due to dynamical friction, and are therefore more likely to be
captured. However, they might not be responsible for the majority of the
binary formation, if the frequency for long-lived transient binaries is
sufficiently small. To address this question we determine the typical time
$t_{Typ}$, required for a transient binary to become bound with the aid of
dynamical friction. We define $t_{Typ}$ as the time it takes for 50\% of all
the KBOs, that form a binary, to become bound for a given strength of
dynamical friction $D$. $t_{Typ}$ is measured from the point at which the
relative separation between the two KBOs is less than $3R_H$. We determine
$t_{Typ}$ in the following way. First, we integrate the same equations as in
\S 4 (i.e. equations \ref{e10} and \ref{e11}). We switch off the dynamical
friction at different times and continue the evolution of the KBOs until $t =
1000 \Omega^{-1}$. This process is repeated until we find the time for which
dynamical friction has to act for 50\% of all KBOs, that form a binary, to
become bound. A transient binary is considered to have become bound when it
remains a binary (i.e. relative separation $ < 3 R_H$) until $t = 1000
\Omega^{-1}$. We repeat this for different $ D $ in order to reveal the
relationship between $t_{Typ}$ and $D$. Again, impact parameters are chosen to
range from $2.2 R_H$ to $3.2 R_H$. Figure \ref{fig3} shows that, for $D
\gtrsim 0.002$, the typical time for permanent capture does not depend
linearly on the strength of the dynamical friction $ D $, but shows a weaker
logarithmic dependence. $t_{Typ} $ only ranges from $ \sim 2 \Omega^{-1} $ for
$ D \sim 0.2$ to $\sim 10 \Omega^{-1} $ for $ D \sim 0.002 $. Furthermore,
figure \ref{fig3} shows a noticeable break around $ D \sim 0.001$; for $D
\lesssim 0.001$, $t_{Typ}$ increase significantly to $ 20 \Omega^{-1} $ and
more. From this, we conclude that longevity of the transient binary (as
discussed by \citet{A05} with $t_{3R_H} \geqslant 15 \Omega^{-1} $) becomes
only important for very weak dynamical friction (i.e. $ D \lesssim 0.002 $)
and is most likely not crucial for KBB formation.  In \S 4 we estimate $D\sim
0.12$, in which case longevity of transient binaries ($ t_{3R_H}\geqslant 15
\Omega^{-1} $) is unlikely to be a major requirement for binary formation.

\section{SUMMARY AND CONCLUSIONS} 
We accurately determine the $ L^2s $ and $ L^3 $ formation rates for
$v<v_H$. We find that while the $L^2s$ formation rate is close to previous
order of magnitude estimates, the $L^3$ formation rate is about a factor of 4
smaller. For $v\ll v_H$, the ratio of the $ L^3 $ to the $ L^2s $ formation
rates is $\sim 0.05(v/v_H)$ and is independent of what sets $D$. It is
therefore independent of the velocity dispersion of the small bodies, their
surface density or the importance of collisions among the small bodies. For
sub-Hill KBO velocities, binaries in the Kuiper Belt formed primarily due to
dynamical friction rather than three body encounters. For super-Hill KBO
velocities ($v \gg v_H$) the $L^2s$ mechanism becomes unimportant. The $L^3$
mechanism forms tight binaries that tend to be saved from break up at a rate
that is reduced by a factor of $(v_H/v)^{11}$ compared to the sub-Hill
case. In addition, the $L^3$ mechanism forms wider binaries $(R_B>R_{crit})$,
at a higher rate that is `only' reduced by a factor of
$(v_H/v)^{6}(v_H/v_B)^5$ relative to the sub-Hill rate. These wide binaries
are constantly created and destroyed leading to an equilibrium abundance of
binaries that scales as $(R_B/R_H)^{3/2}$. Whether and how any of these wide
binaries would survive the dynamical excitation of the Kuiper Belt remains an
open question.

In addition, we determine the frequency of long-lived transient binaries. We
show that the frequency of long-lived transient binaries drops exponentially
with the system's lifetime for $v_{rel}<v_H$. About 1000 transient binaries
occur with $ t_{3R_H} \gtrsim 3 \Omega ^{-1} $ for each transient-binary with
$ t_{3R_H} \gtrsim 15 \Omega ^{-1} $. The long-lived transient binaries
investigated by \citet{A05} and \citet{A07} are therefore very
rare. Long-lived transient binaries are not important for binary formation via
the $L^3$ mechanism, since the binary formation rate scales only linearly with
transient-binary lifetime, but the frequency of transient binaries drops
exponentially as a function of its lifetime. Long-lived transient binaries do
not exist for $v_{rel}\gg v_H$.  We show that the apparent shortage of
binaries forming from short-lived transient binaries (i.e. $t_{3R_H}\lesssim
2.5 \Omega^{-1}$) found by \citet{A07} can be explained by a bias in their
initial conditions that discriminates against binary formation from transient
binaries with $t_{3R_H}\lesssim 5 \Omega^{-1}$.  Finally, to assess the
importance of long-lived transient binaries in the $L^2s$ mechanism, we
determine the typical time $ t_{Typ} $ required for a transient binary to
become bound with the aid of dynamical friction. We show that longevity of
the transient binary (as discussed by \citet {A05} with $ t_{3R_H} \geqslant
15 \Omega^{-1}$) only becomes important for very weak dynamical friction
(i.e. $D \lesssim 0.002$). We estimate $D\sim 0.12$, in which case longevity
of transient binaries ($ t_{3R_H}\geqslant 15 \Omega^{-1} $) is unlikely to be
a major requirement for binary formation.

\acknowledgments We thank Peter Goldreich for stimulating discussions and the
anonymous referee for valuable comments that helped to clarify the
manuscript. Some of the numerical calculations presented here were performed
on Caltech's Division of Geological and Planetary Sciences Dell
cluster. R. S. is an Alfred P. Sloan Fellow and a Packard Fellow.

\clearpage 

\end{document}